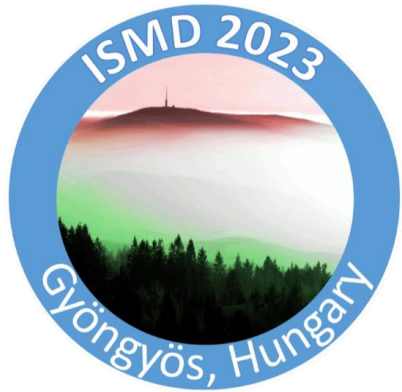

# A new game with Quark Matter Cards:

# Interactions of elementary particles

Ana Uzelac[1]

[1]Zagreb, Croatia

December 27, 2023

**Abstract**

This paper introduces a card game called "Quark Matter Card Games," inspired by the creativity of a high school student, Csaba Török, and developed in collaboration with physicist Tamás Csörgő. The game utilizes a deck of 66 cards representing elementary particles and antiparticles, offering an entertaining way to popularize science and introduce players to particle physics concepts. The initial edition includes games exploring topics like baryon formation, mesons, quark colors, and more.

The author proposes a new game focusing on the four fundamental forces. Players must strategically place cards on a central card, simulating interactions based on strong, electromagnetic, weak, and gravitational forces. Specific rules for particle placement, including fundamental forces, annihilation and neutrino oscillations are elucidated. Additional rules and conditions add complexity and strategy to the game, ensuring active engagement.

The intended audience ranges from individuals familiar with particle physics to those new to the field. The game provides an engaging platform for learning about elementary particle interactions, with varying levels of complexity. The paper discusses the educational potential, offering suggestions for simplified versions. Furthermore, all necessary concepts are briefly explained, and the physical background of the game is provided.

The paper concludes with topics for further discussions, linking game experiences to particle physics concepts. Questions cover gravitational interaction interpretations, differentiation between quarks and leptons, explanations of weak and electromagnetic interactions and more. The acknowledgment section expresses gratitude to mentors and pioneers of card games with elementary particles for their inspiration and support.

# 1   Introduction

Csaba Török, a seventeen-year-old high school student in Hungary, on New Year's Eve 2008/2009, inspired by discussions about the physics of elementary particles, conceived the idea of creating a deck of cards depicting elementary particles, from which he developed the game "ANTI."

This game was later perfected and expanded, leading to the creation of a deck of cards with a set of games called „Quark Matter, Card Games" whose primary goal was entertainment, and secondary goals included the popularization of science and introducing players to basic concepts of particle physics. During the development of this game, Csaba teamed up with his peer Judit Csörgő. Tamás Csörgő, Judit's father and, as well as a research physicist engaged in experimental and theoretical high-energy physics related to the RHIC accelerator at Brookhaven, joined the team as a mentor and manager.

In the initial edition of *Quark Matter Card Games*, four different games are described that can be played using a deck of 66 cards representing some of the elementary particles (as well as antiparticles) of the Standard Model. [1][2]

In the deck of cards from [2] , there are 66 cards, with each card representing a particle or antiparticle. Quarks and antiquarks are distinguished by colors, while leptons and antileptons are indicated by black-and-white cards.

The quarks represented on the cards are the up (*u*), down (*d*), and strange (*s*) quarks, and each of them can be colored in red, green, or blue, signifying the color charge of the individual quark. The leptons represented on the cards in the deck are leptons from the first and second generations, namely, the electron ($e^-$), electronic neutrino ($\nu_e$), muon ($\mu^-$), and muonic neutrino ($\nu_\mu$). Additionally, the deck includes antileptons corresponding to these leptons, which are marked with plusses next to the particle names, as shown below[1]:

---

[1] Pictures have been made using templates from [2]

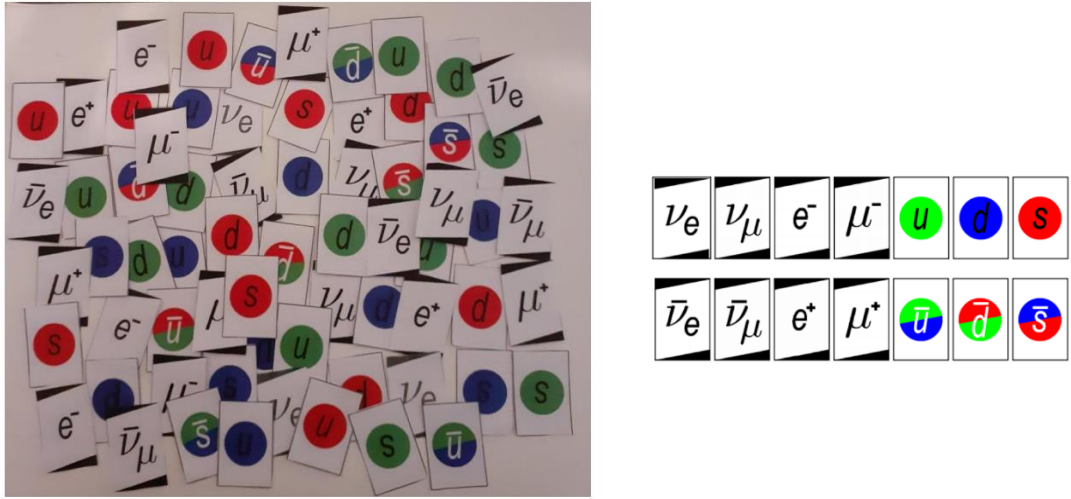

Figure 1 - Examples of different cards from the deck

Since then, several new games have been devised using the same deck of cards.

In these games, players with different levels of physics knowledge can not only have fun but also acquaint themselves with or enhance their understanding of various fundamental concepts in elementary particle physics. These include the names of elementary particles and antiparticles, the formation of baryons and mesons, the colors of quarks and hadrons, the electric charge of elementary particles, as well as baryon and lepton numbers and their conservation.[2]

Moreover, through such card games, players (individually or as part of a workshop, school, or science club) can delve into a wide range of additional topics. For instance, topics related to antiparticles and annihilation, the physics of the early universe, processes in accelerators and particle colliders, the so-called quark matter and quark-gluon plasma, particle decays and conservation laws, the phenomenon of cosmic rays, and the Higgs boson and experiments leading to its discovery.[2]. From the original deck of cards, entirely new and interesting ideas for games have emerged, such as Particle Poker[3] and the Rubik's Cube of Quark Matter.[1][4].

Due to the numerous interconnected themes in these games, they can serve as an excellent source for introducing (through entertainment) individuals to these seemingly complex and challenging areas of physics.

In my last year of studying physics education, inspired by all of these games that use a deck of cards with elementary particles, I decided to try my hand at designing new games with this deck, which could be related to aspects of elementary particle physics that were not already mentioned in existing games.

So I came up with an idea of a game in which four fundamental forces could be mentioned. In the continuation of this article, the mentioned game will be presented and explained.

The game is intended for **two to four players** (due to the limited numbers of particle and antiparticle cards in the deck).

The basic idea is that the rules of card placement correlate with elementary particle interactions for each of the four fundamental forces.

## 2   The beginning of the game

At the beginning of the game, players need to choose **one card** from the deck. That card must must represent a **particle**, not an antiparticle (so, for example, it could be red coloured *s*-quark). The chosen card must be places face-up in the center of the table.

The remaining 65 cards should be thoroughly shuffled and then placed in a circle around the central card, with these cards in the circle facing downwards, as shown in the following image:

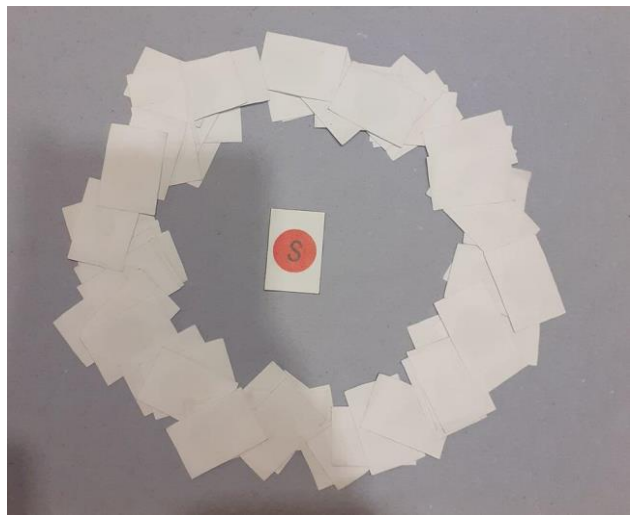

Figure 2 - An example of card placement at the beginning of the game. In this case, the central card is the red s-quark.

Then, each player **draws four cards** from the circle of cards placed around the central card. The drawn cards are held by the players in their hands, so only they can see which cards they have.

At this point, it is necessary to determine the order in which players will take their turns (for example, in a clockwise direction). Then, the gameplay begins! Players must take turns placing one card of their choice (from the cards they hold in their hands) on the central card, while naming the interaction according to the following rules.

# 3 Rules of card placement: Placing particles on the central card

As it was previously mentioned, the rules for placing cards on the central card correlate with the conditions of interactions between elementary particles for each of the four fundamental forces. Specifically, **it is only possible to place a card on the central card if it represents a particle that can interact with the central card, and it is necessary to always state the type of interaction**![5]

Accordingly, the following four interactions are possible.

<u>STRONG INTERACTION:</u>

Due to the short range of the strong force, it is assumed in this game that **only quarks within the same baryon can interact strongly**. During this interaction, there is an exchange of color between them, while maintaining the overall color neutrality of the baryon.

As a result, the following moves are allowed: **for each quark ($u$, $c$, or $d$), you can place any other quark ($u$, $c$, or $d$), provided that their color is different from the initial quark's color**! For example, on the red $s$-quark from Figure *2 - An example of card placement at the beginning of the game. In this case, the central card is the red s-quark.* previous figure, you can place a blue or green $u$, $s$, or $d$ quark! To make the move valid, it is necessary to state explicitly that it is a strong force interaction.

<u>ELECTROMAGNETIC INTERACTION:</u>

The fundamental property of electromagnetic force is that only two electrically charged particles can interact with it. Accordingly, you can place **any electrically charged particle ($u$, $s$, or $d$ quark of any color, $e^-$ or $\mu^-$) on any other electrically charged particle** (i.e., any card representing $u$, $s$, or $d$ quark, $e^-$ or $\mu^-$). Once again, it is crucial to clearly state that it is an electromagnetic interaction.

Taking the example of the central card from Figure , players can place any of the following cards on the red *s*-quark: any *u*, *s*, or *d* quark, *e*[-] or *μ*[-]; basically any card except the electrically neutral electron and muon neutrinos.

WEAK INTERACTION:

In this type of interaction, particles that can interact are leptons of the same generation and certain combinations of quarks (within or outside the same generation). More about specific possible transitions can be found in Chapter 7 of this paper.

Hence, the rules are as follows:

| LEPTONS | QUARKS |
|---|---|
| • An **electron** can be placed on an **electronic neutrino**, and vice versa.<br><br>• **A muon can be placed on a muonic neutrino, and vice versa.** | • An "*u*" quark can have a "*d*" quark or "*s*" quark placed on it, but only if they are of **the same color** as the "*u*" quark.<br>• A "*d*" quark can have only an "*u*" quark of the **same color** placed on it.<br>• An "*s*" quark can have only an "*u*" quark of **the same color** placed on it. |

*Table 1 - Rules for card placement according to weak interaction*

The fact that quarks placed on the given quark must be of the same color reflects the observation that **the weak force can change the flavor but not the color of the quark**.

So, if we take the example that the central card is a red *s*-quark, according to the weak force interaction, we can only place a red *u*-quark on it! Of course, in this case it is necessary to specify and say that it is a weak interaction![5]

GRAVITATIONAL INTERACTION:

Considering that gravitational interaction can occur between any two particles with mass, if we were to follow the stacking rules for the gravitational force based on the same principles as before, it would imply that any card from the deck could be placed on any central card. However, gravitational interaction between elementary particles is extremely weak due to their small masses, and this is reflected in the game as follows:

**Only once during the game** (but not when the player has only one card left in hand, and all the cards from the central circle have been taken), the player can take advantage of the opportunity for "gravitational interaction" and **place any particle card** on the central one.

This situation can be very intriguing as it may change the course of the game and give a significant advantage to the player who utilizes it at the right moment.

In this case, the card placement rule related to gravitational force can teach us that it is universal, but extremely weak when masses are low.

*ADDITIONAL RULE (for additional knowledge and more fun) – NEUTRINO OSCILLATIONS*

If there is **an electronic neutrino** at the center of the table, **it is possible to place a muonic neutrino on it**, and vice versa. At that moment, it is necessary to state that it is a neutrino oscillation!

With this addition, players can become acquainted with the discovery of "neutrino oscillations," the periodic changes of neutrinos from one type to another, indicating that neutrinos have nonzero rest mass. Takaaki Kajita and Arthur B. McDonald were awarded the Nobel Prize in Physics in 2015 for this discovery.[6]

# 4 Rules of card placement: Placing antiparticles on the central card

Since the deck also has antiparticle cards , the rule of an annihilation has been designed– for instance, if there's a muon on the center of the table, an antimuon can be placed on it.

In this game, **cards representing antiparticles can be placed on the central card only if it truly represents the antiparticle of the central card**! In that situation, the player must say "Annihilation!", which corresponds to the physical phenomenon of matter-antimatter annihilation. Therefore, the following moves are possible:

| *LEPTONS* | *QUARKS* |
|---|---|
| <ul><li>A positron can be placed on an electron.</li><li>An electron- antineutrino can be placed on an electron-neutrino.</li><li>An antimuon can be placed on a muon.</li><li>A muon- antineutrino can be placed on a muon neutrino.</li></ul> | <ul><li>$\bar{u}$ (antiquark of u) can be placed on a u quark, but only if the antiquark is of the corresponding anti-color. For example, on a blue u quark, you can place only an anti-blue u antiquark (i.e., a u card colored red-green).</li><li>Accordingly, $\bar{s}$ (antiquark of s) and $\bar{d}$ (antiquark of d) can be placed on s and d quarks, respectively.</li></ul> |

Table 2- Rules for placing antiparticles on a central card

Since the probability of that move is very low, this move is more *"valuable"* – so if a player notices it and performs it, it gains an advantage!

The placement of corresponding cards representing antiparticles on the central card is highly useful and desirable in the game because, after achieving an "Annihilation," the player can, in the same move, **place any of their other cards on the table**, thus getting rid of two of their cards in one move. Additionally, by placing a card of their choice in the central row, the player **intentionally changes the course of the game**, allowing for strategic moves. For instance, by placing a neutrino, there's a high chance that the next player may face difficulties since neutrinos do not interact with a large number of available particles.

# 5 Additional rules and game ending

In order to make the game even more interesting and dynamic, five additional rules have been designed. Here they are:

I. If a player fails to state the appropriate interaction during their turn, and other player(s) remind them, the player must draw **three additional cards** into their hand (unless there are no more available cards in the central pile).

II. If a situation arises where a player places a card on the central card that can interact with it through more than one fundamental interaction (excluding gravitational), and the player only states one of them, any other player can "correct" them by specifying the exact type of interaction the player forgot to mention. In this case, the player who forgot to state all possible interactions must draw **two additional cards** from the center of the table.

For example, if a player places a blue *u*-quark on a red *s*-quark in the center and only mentions electromagnetic interaction, another player can correct them by stating that a strong interaction is also possible between these two particles, thus forcing them to draw two additional cards.

III. If a situation occurs where a player incorrectly "corrects" another teammate, they themselves **must draw two cards** from the center of the table as a penalty.

This way, the game becomes more interesting and prevents players from simply stating electromagnetic interaction between electrically charged particles without considering other possible interactions. Therefore, players must always be attentive to the moves of others and actively think about possible interactions.

IV. If a situation arises where the player whose turn it is does not have a suitable card in their hand that they could (or know how to) place on the central card and name the corresponding interaction, they must draw cards from the center pile until they get a card they can place on the central card. If such a situation occurs to a player when all cards from the central circle have already been taken, the player must draw cards in order from those that have already been placed on the central card, starting from the bottom of the pile.

V. If during the game any of the players run out of cards in their hands, but not all cards from the central pile have been taken, then the player who is out of cards must **randomly draw cards from the central pile** until they draw a card they can place on the central card, just like in the previous situation.

The game ends when all cards from the central pile have been taken, and one of the players has **run out of all cards representing particles** (they are allowed to have cards representing antiparticles in their hands!).

That player is the winner, and the other players are ranked according to the number of cards representing particles they have left in their hands, with the player holding the most cards representing particles being ranked last.[5]

# 6 For whom is the game intended?

Due to the fact that the game relies on the fundamental interactions of elementary particles, individuals who are familiar with the basics of particle physics and the four fundamental interactions will quickly grasp the rules of the game. In this case, it is sufficient to know that only particles capable of interacting with the central card should be placed on it, and the player must state the type of interaction. With this knowledge, the game can immediately begin.

Such players might find this game particularly interesting due to its connection to physics, as well as the possibility to strategize by placing particles that do not interact with many others in the right situations. For example, utilizing moves like "annihilation" or "gravitational interaction" at the right moment can allow players to place a card that poorly interacts with others (like a neutrino) as the central card or predict which card the opponent may not have based on their previous moves.

Moreover, individuals well-versed in particle interactions can quickly notice mistakes made by others, such as when teammates fail to mention correct or all possible interactions. By recognizing such situations, players can force their opponents to draw additional cards and gain an advantage over them.

However, this game can still be played by individuals who are not well acquainted with elementary particle interactions (and by playing, they can greatly increase their knowledge). In such cases, one can start by learning the rules of the game, which can later be supplemented with the physics background. The rules of the game are not any more complicated than typical card games, and mastering them automatically means that the person has learned which particles can interact through specific interactions!

In my opinion, even children in middle school (with some simplifications) who are learning about elementary particles and fundamental forces can easily master this game. In this scenario, the game can be an excellent tool for active consolidation and deepening of knowledge. For example, with younger children, simpler versions of the game can be introduced, where only cards representing particles are used, and rules of card placement are introduced gradually, paralleled with the introduction and learning of specific forces. Later on, antiparticles and related rules can be introduced once their properties and differences compared to particles are understood.

Through this game, children can develop their understanding of physics and enhance their logical thinking. It can stimulate their curiosity and interest in science, encouraging them to further educate themselves in the field of physics and natural sciences.

## 7   Physical background of the game: About elementary particle interactions

In addition to the rules of the game, which are correlated with various types of particle interactions, players must be familiar with basic concepts related to elementary particle physics. Specifically, they need to know of the elementary particles and antiparticles represented by the cards in the deck, as well as their electric charges. For quarks and antiquarks, they must understand the concept of "color charge" and the overall color neutrality of hadrons.

In order for this article to be useful for those who independently want to learn all of the above, and much more (and do not have a broad knowledge of physical concepts), in this chapter these concepts are listed and briefly explained.

### Six quarks and six leptons

According to current knowledge, all elementary particles can be divided into three fundamental categories - quarks, leptons, and gauge bosons.

Leptons are elementary particles that, unlike quarks, **do not** interact via the strong interaction. All leptons have a spin of 1/2, which places them in the category of fermions. The name "lepton" comes from the Greek word "λεπτός" (leptós), which can be translated as "small, tiny, thin." Physicist Léon Rosenfeld introduced this term in 1948 to refer to the then-known elementary particles with the smallest masses[2], specifically particles with masses significantly smaller than the mass of the atomic nucleus.

There are **six different types, or "flavors," of leptons**. Of the six existing leptons, three are electrically charged, and three are electrically neutral. Charged and neutral leptons can be grouped into so-called "generations," resulting in the electron, muon, and tau generations. For each lepton, there is also its antiparticle (or antilepton) whose properties are equivalent to the comparison of earlier properties between positrons and electrons, as well as antiprotons and protons.

The list of leptons and their corresponding antiparticles, along with their symbols and masses, are provided in the following table[3]:

| Lepton name | Symbol | Antiparticle | Mass (MeV/$c^2$) |
|---|---|---|---|
| Electron | $e^-$ | $e^+$, | 0,511 |
| Electron neutrino | $\nu_e$ | $\bar{\nu}_e$ | $< 3 \cdot 10^{-6}$ |
| Muon | $\mu^-$ | $\mu^+$ | 105,7 |
| Muon neutrino | $\nu_\mu$ | $\bar{\nu}_\mu$ | $< 0,19$ |
| Tau | $\tau^-$ | $\tau^+$ | 1777 |
| Tau-neutrino | $\nu_\tau$ | $\bar{\nu}_\tau$ | $< 18,2$ |

*Table 3 - Six different leptons*

Electrons, muons, and tau leptons are negatively charged (indicated by a minus sign in their symbols), while electron neutrinos, muon neutrinos, and tau neutrinos are electrically neutral particles.[7]

---

[3] The table was made using data from [7]

Unlike leptons, quarks are elementary particles that **can** interact through the strong force! The number of different quarks is the same as the number of different leptons, so there are **six different flavors of quarks**, which can also be divided into three generations. Quarks possess a property called **"color charge,"** which is why (unlike leptons) they always combine in groups in nature; such as baryons and mesons.

The term "quark" was introduced by Murray Gell-Mann, based on a quote from James Joyce's novel "Finnegans Wake." Regarding this event, Gell-Mann said in his book "The Quark and the Jaguar: Adventures in the Simple and the Complex":

*"In 1963, when I assigned the name 'quark' to the fundamental constituents of the nucleon, I had the sound first, without the spelling, which could have been 'kwork.' Then, in one of my occasional perusals of Finnegans Wake, by James Joyce, I came across the word 'quark' in the phrase 'Three quarks for Muster Mark.' ... In any case, the number three fitted perfectly the way quarks occur in nature."*

Namely, each baryon represents a bound state of three valence quarks, each antibaryon is a bound state of three antiquarks, and mesons are always bound states of quarks and antiquarks.

This statement can be illustrated with the following diagram:

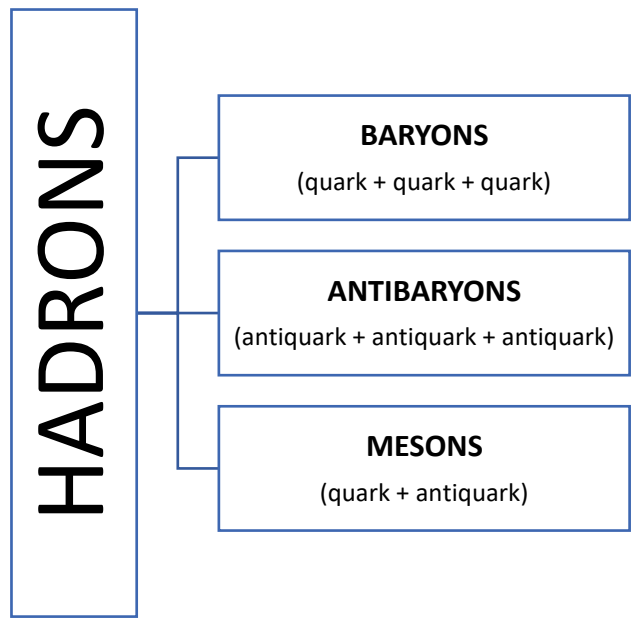

Figure 3 – Hadrons and their quark matter

The beginnings of the theory of quarks in 1964 predicted the existence of three different quarks, which were named up, down, and strange quark. It was found that these quarks make up protons, neutrons, π and K-mesons, as well as many other known hadrons. For example, a proton consists of two up quarks and one down quark, while a neutron consists of two down quarks and one up quark.[7]

An interesting consequence of this is the fact that quarks must be particles with **electric charge values that are not integer multiples of the elementary charge!** For instance, considering the quarks that make up protons and neutrons, we can deduce that the electric charge of the up quark must be +2/3 *e*, and the charge of the down quark -1/3 *e*, where *e* represents the value of the elementary charge (which is equivalent to the electric charge of one electron or muon).

The principle of determining these electric charge values is explained in the following diagram:

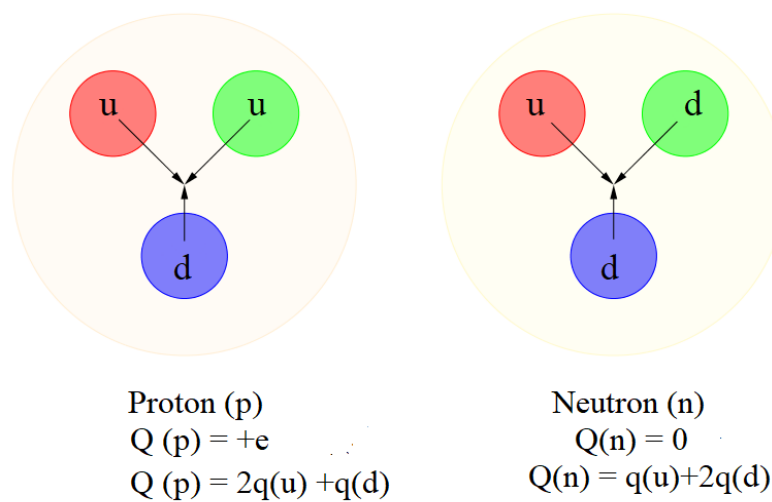

Figure 4 – *The principle of determining the electric charge of the upper (u) and lower (d) quarks from the known charges of protons and neutrons*

As visible in Figure 4, the colored circles and corresponding labels represent the quarks that make up the proton (on the left) and the neutron (on the right). Since the total electric charge of the proton (Q(p)) and neutron (Q(n)) is known, it is easy to determine the charge of the up and down quarks (denoted as q(u) and q(d)) by setting up equations as shown in the figure, and then solving the system of these two equations, where the charges of the quarks are the unknowns.

In 1974, the existence of a fourth "charm" (c) quark was confirmed by studying ψ mesons at Brookhaven, and in 1977, a new meson - ϒ, and its constituent, the fifth quark, were discovered, named the "bottom" (b) quark. Along with the then-known six leptons, many scientists also considered the possibility of a sixth quark, which was named the "top" (t) quark. In 1995, experiments at Fermilab's Tevatron indeed discovered such a quark, whose properties matched the predictions.[7]

Therefore, there are six different types (flavors) of quarks, all of which, along with their main properties, are listed in the following table:

| Name | Symbol | Charge $Q$ (e) | Mass | Strangeness $S$ | Charm $C$ | Bottom $B$ | Top $T$ |
|---|---|---|---|---|---|---|---|
| *Up* | u | $+\frac{2}{3}$ | $\cong 2{,}2 \text{ MeV}/c^2$ | 0 | 0 | 0 | 0 |
| *Down* | d | $-\frac{1}{3}$ | $\cong 4{,}7 \text{ MeV}/c^2$ | 0 | 0 | 0 | 0 |
| *Strange* | s | $-\frac{1}{3}$ | $\cong 96 \text{ MeV}/c^2$ | −1 | 0 | 0 | 0 |
| *Charm* | c | $+\frac{2}{3}$ | $\cong 1{,}28 \text{ GeV}/c^2$ | 0 | +1 | 0 | 0 |
| *Bottom* | b | $-\frac{1}{3}$ | $\cong 4{,}18 \text{ GeV}/c^2$ | 0 | 0 | +1 | 0 |
| *Top* | t | $+\frac{2}{3}$ | $\cong 173{,}1 \text{ GeV}/c^2$ | 0 | 0 | 0 | +1 |

Table 4 – Six flavours of quarks

*Note:* The letters S, C, B, and T denote new quantum numbers that describe different properties of the mentioned quarks and the hadrons they form. Data on quark masses are taken from [8], and other data are taken from [7].

For instance, certain hadrons composed of strange (s) quarks (such as kaons, lambda, and sigma mesons) were observed to possess an unusual property - they could be easily produced in particle collisions, but they decayed much more slowly than expected.

Due to the fact that such mesons are always produced in collisions in pairs, a new quantum number was postulated, called "Strangeness" (S), which is conserved in the creation but not in the decay of hadrons containing s quarks. According to this idea, the strangeness of an individual hadron is equal to the total sum of strangeness of the quarks it consists of. Strangeness is conserved in strong and electromagnetic interactions but not in weak interactions! [9]

Just like with leptons, there is also an corresponding antiparticle for each quark - the antiquark. They are denoted by a dash above the corresponding symbol, so the following antiquarks exist: ū, d̄, s̄, c̄, b̄, and t̄. For antiquarks, the values of Q, S, C, B, and T listed in the table are opposite to those of their corresponding quarks, while their masses are the same.

Colors of Quarks and Formation of Hadrons

Since quarks, like leptons, belong to the group of fermions, they are subject to the Pauli exclusion principle, which would make the existence of baryons consisting of two or three identical quarks impossible. Consequently, neutrons and protons, which make up atomic nuclei, could not exist![5]

To resolve this dilemma, it was postulated that quarks of the same flavor can appear in **three different variations**, which were called **"colors"** of quarks. Therefore, there are three different "colors" of quarks - red, green, and blue, and for example, a blue up quark differs in properties from a red up quark. In this way, the Pauli exclusion principle is not violated if we assume that inside a hadron, there can be quarks of the same flavor but different colors. In contrast to quarks, their antiquarks are assigned the property of anti-colors. For example, the antiquark of a red down quark is an anti-red down antiquark, which can be described as green-blue or "non-red" color.

It is important to note that the colors of quarks have no connection with "optical" colors, which physically correspond to the wavelengths of electromagnetic radiation. Instead, the term "color" in this case describes a specific property (which cannot be numerically described) that appears in three different variations for each flavor of quark. That is why this property is sometimes called "color charge."

The term "color" was nevertheless introduced because in nature, quarks always combine to form hadrons (baryons, antibaryons, or mesons) in a very specific way. **Baryons must contain three quarks, each of a different color, and their antiparticles, antibaryons, must be composed of antiquarks of different anti-colors. Mesons, which are composed of a quark and an antiquark, must be formed from a quark of one color and an antiquark that corresponds to the anti-color of that quark.** For example, a valid meson is one composed of any red quark and any anti-red antiquark. Therefore, the formation of colored quarks into larger particles can be compared to "mixing" the mentioned colors (red, green, and blue) so that the resulting product (hadron) is color-neutral (meaning it lacks the property called "color charge").

For example, by mixing three different colors of light - red, green, and blue, a multitude of shades of different colors can be produced, as shown in figure below. This method is used to create colored images in most modern displays and screens, such as those in mobile phones and computer screens. When equal intensities of red, blue, and green colors are mixed in equal proportions, a "white" color is produced.

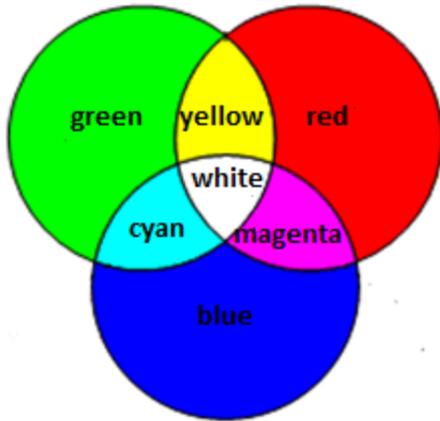

The additive mixing of red, green, and blue colors, which produces white color, can be compared to the property of color neutrality in baryons: **Every baryon must always contain one red, one green, and one blue quark, so it can be said that the baryon as a whole has a "white" color** (meaning it lacks the property of "color").

*Figure 5 – The additive mixing of red, green and blue colors*

As can be observed in Figure 5, the "cyan" color is formed by mixing green and blue colors, and it lacks the red color, which is why this color can also be called "anti-red." Similarly, the yellow color can be considered "anti-blue," and magenta as "anti-green." By mixing these "anti-colors," black color is obtained, as shown below

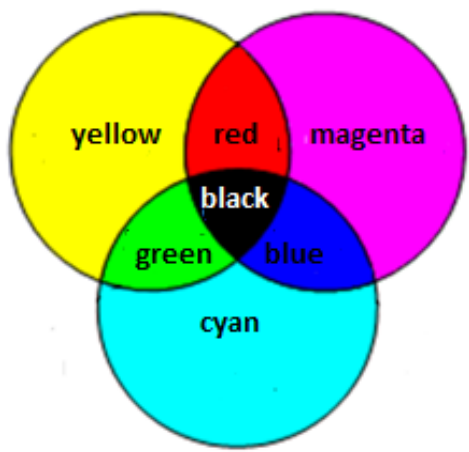

By mixing "anti-blue," "anti-green," and "anti-red" colors in equal proportions, black color is formed – which can also be called "anti-white."

Similarly, anti-baryons always consist of one anti-red, one anti-green, and one anti-blue quark, which is why they are ultimately black or "anti-white" – devoid of the property of "color."[5]

*Figure 6 – The additive mixing of „anti" colors*

By combining a color with its own "anti-color" in equal intensities and proportions, white color is always obtained, as shown in Figure 7.

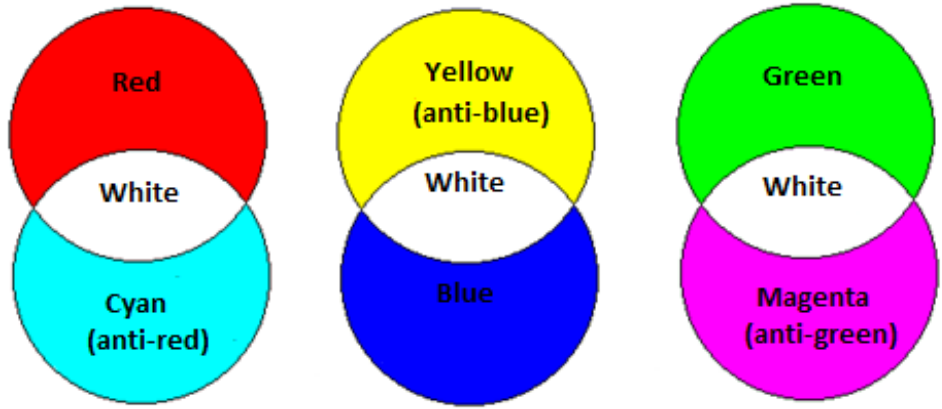

*Figure 7 – Possible mixtures of color – anticolor pairs*

Therefore, this image also represents the possible color combinations of quark-antiquark pairs in the formation of mesons. Mesons are composed of any quark and antiquark that have corresponding equal colors and anti-colors, so it can be said that as a whole, they do not have the property of color.

The fact that hadrons always exist in a state of "no color" can be summarized as follows:

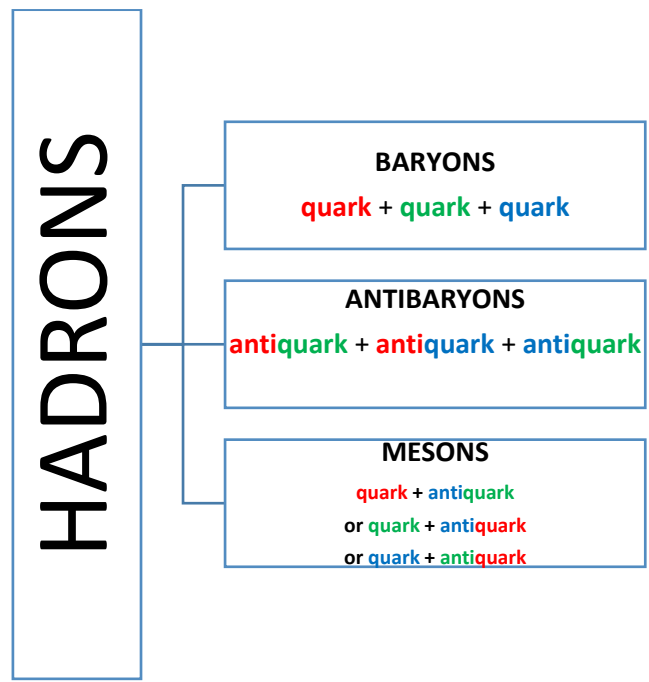

*Figure 8 – Hadrons according to color neutrality*

The terms "quark" and "antiquark" designate quarks and antiquarks of any flavor with the property of color charge equal to the color mentioned in the text. For example, the expression "quark + antiquark" indicates that in the case of a valid meson, it must consist of any red quark and any blue-green (or antired) antiquark.

The theory of strong interactions, in which quarks (and other "colored" particles - gluons) interact, relies on this color charge property. Due to the analogy with colors, this theory is called Quantum Chromodynamics (QCD). Precisely because quarks in hadrons (baryons and mesons) always appear in bound states, i.e., in colorless groups, QCD predicts that it is impossible to isolate individual quarks in independent, free states similar to how leptons can exist.[5]

## Interaction of Elementary Particles - Four Fundamental Forces and Exchange Particles

According to the current understanding of physics, every interaction between elementary particles (and thus between different bodies) can be described using **four fundamental forces** (interactions) - gravitational, electromagnetic, strong, and weak force.[7]

In the Standard Model, each of these interactions is mediated by the exchange of corresponding particles, which we call **exchange particles** or **force carriers**. Below, we will provide an overview and a brief explanation of each of these forces and their exchange particles at the level of elementary particles. The forces will be described according to their relative strength from weakest to strongest, and for weak, electromagnetic, and strong interactions, a concrete example of the interaction of elementary particles using Feynman diagrams will be given.

### Gravitational force

Due to the fact that gravitational interaction is **the weakest** (about $10^{-40}$) times weaker in relative strength compared to the strongest force - strong force!)[10], its influence becomes noticeable only among bodies composed of an extremely large number of particles, such as planets, stars, and other celestial bodies. When considering elementary particles, its effect is so tiny that it can be neglected.

Nevertheless, gravitational interaction exists between all mass elementary particles, and its range is infinite, although the strength of this force decreases with the square of the distance between particles.

Gravitational interaction is always **attractive**, and the predicted exchange particle is called the **graviton.** The predicted rest mass of the graviton is zero, and its spin is 2, making it stable particle. However, due to the weak strength of gravitational force between elementary particles, the graviton has not been experimentally observed in laboratories yet.[11]

## Weak force

Although the weak force is named "weak," it is **not actually the weakest** of all fundamental forces. Its relative strength is $10^{-5}$ times weaker than the strong force, but it is still $10^{35}$ times stronger than the weakest force, the gravitational force.[10]

**Both quarks and leptons** can interact with this interaction, and the exchange particles for the weak force are called weak or intermediate bosons. There are three types of these bosons: $Z^0$, $W^+$, and $W^-$. As indicated by the subscripts of their symbols, $Z^0$ bosons are electrically neutral, $W^+$ bosons are positively charged, and $W^-$ bosons are negatively charged. All of them are unstable particles. Their existence was predicted in 1957 and confirmed in 1983 through experiments at CERN, where protons and antiprotons collided. For their work, scientists Carlo Rubbia and Simon Van Der Meer were awarded the Nobel Prize in 1984.[7][10]

All these bosons have a spin of 1 and are not massless. The rest masses of the $W^+$ bosons are 80.4 GeV/$c^2$, and the rest mass of the $Z^0$ boson is 91.2 GeV/$c^2$.[7] Due to the significant masses of these bosons, the weak force has a very short range, about 0.001 fm, which is roughly a hundred times smaller than the typical diameter of an atomic nucleus.[7][5] Because of its limited range, the weak force is often referred to as the weak nuclear force.

At the level of elementary particles, we can say that the weak nuclear force acts by **"changing" elementary particles from one type to another**.

For leptons, the rule is that the weak force can only act **between leptons within the same generation**, but not between generations.[12]

For example, through the action of the weak force, specifically the emission of a $W^-$ intermediate boson, a muon can transform into a muon neutrino (both have the same lepton number of 1), an electron, and an electronic antineutrino (which results from the $W^-$ decay and does not change the net lepton number).[5]

Similarly, the decay of a tau lepton can be explained by its transformation through the emission of a $W^-$ boson into a neutrino of the same generation, namely the tau-neutrino.

Again, the intermediate W⁻ boson decays into possible products, such as electron-antineutrino or muon-antineutrino pairs.

Possible lepton transitions are illustrated in the image below:

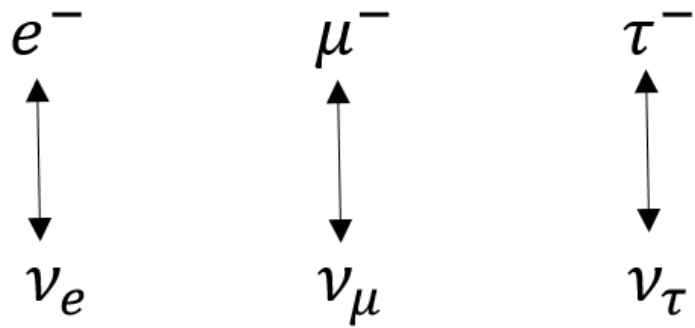

*Figure 9 – Possible weak force lepton transitions*

*Note:* Figure 9 schematically represents the leptons that interact through the weak force; specifically, those leptons that must necessarily converge at the same vertex of Feynman diagrams connected to the corresponding intermediate bosons. However, in the case of the previously mentioned decays of heavier leptons, other products resulting from the decays of intermediate bosons must not be overlooked.

Regarding quarks, through the action of the weak nuclear force, **quarks can transition from one type to another within the same generation as well as between**

**generations**. Possible transitions of quarks from one to another through the action of the weak force are illustrated in the following image:

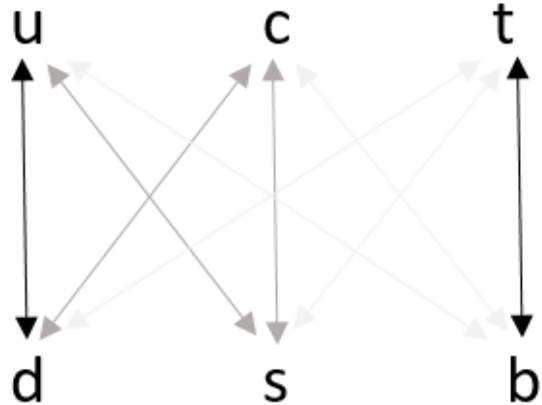

*Figure 10 – Possible weak force quark transitions[4]*

The color of the arrows corresponds to the transition probabilities of the given quarks, with the darkest arrows representing the most common transitions, and the lightest arrows representing the rarest transitions. (These transition probabilities are given by the Cabibbo-Kobayashi-Maskawa matrix[13][5])

It is important to note that during such flavor-changing transitions of quarks, the color of the quarks remains the same as before! (The change of quark color is governed by another fundamental force - the strong force, which will be discussed later).

The weak interaction is responsible for the decays of unstable composite particles (such as mesons) and also plays a crucial role in nuclear fusion processes inside stars. For example,

---

[4] Figure is made with data from [12] and [13]

it leads to β⁻ decay, as depicted in the following Feynman diagram[5]:

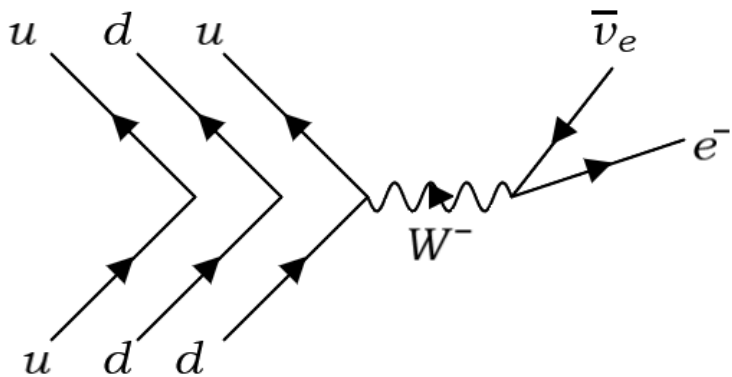

Figure 11 - Feynman diagram of beta – minus decay

So, in the neutron, one of the down (*d*) quarks transforms into an up (*u*) quark, emitting a virtual W⁻ boson, while the up quark absorbs that boson, transforming back into a down quark. Since the W⁻ boson also interacts with leptons while conserving lepton number, it decays into an electron and its neutrino. (Muon and tau leptons are much heavier, and energy conservation forbids their production due to the small mass difference between the neutron and the proton). By initially changing the down quark into an up quark, the proton transformed into a neutron. The color of the up quark remained the same as the color of the initial down quark because in baryons, the total color must be "white," meaning that all three existing quark colors must be present.[5]

Electromagnetic force

This force, like the gravitational force, is well known in classical physics, and its strength also decreases with the square of the distance. Like the gravitational force, the range of the electromagnetic force is practically infinite, which requires the exchange particle for this force to be massless as well.

The exchange particle for the electromagnetic force is the **photon**, which has a spin of 2. Unlike the gravitational force, the electromagnetic force can be **attractive or repulsive**. Particles with the opposite type of charge (positive and negative) attract, and particles with the same type of charge repel.

The electromagnetic force can appear with all particles that are electrically charged – so all **charged** quarks and leptons, but not electrically neutral neutrinos can interact with it. During electromagnetic interactions, electrically charged particles interact with each other by **emitting virtual photons** and simultaneously **absorbing photons** emitted by the other particle. As an example of such interaction, the Feynman diagram below shows electromagnetic interaction of two electrons.

---

[5] Every Feynman diagram in this paper has been created with [14]

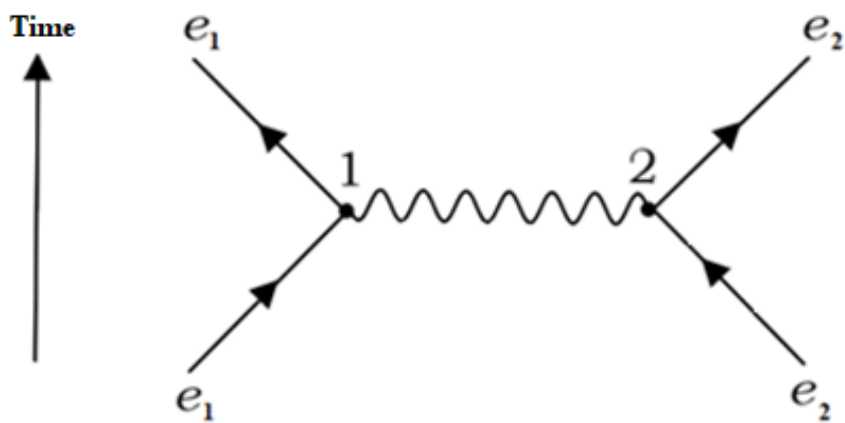

Figure 12- Feynman diagram of electromagnetic force in case of two electrons

So, at point 1, photon $e_1$ emits a virtual photon, represented by the wavy curve, which is absorbed by electron $e_2$ at point 2.

Strong force

**Only quarks can interact with strong force**. That is the force that holds nucleons in the atomic nuclei together, overcoming the repulsive effect of the electromagnetic force. At a distance of 1 fm, the strength of the strong force is approximately 137 times greater than the strength of the electromagnetic force.[7]

The fact that the strong nuclear force acts only on quarks, but not on leptons, is one of the most important differences between these two groups of fermions, and it explains why all quarks in the universe are grouped together in color-neutral combinations, while this is not the case for leptons.

When two quarks interact through the strong force, they **exchange** particles called **gluons**. Gluons, like quarks, are "colored" particles, so during the gluon exchange, the color of the quarks changes. Since gluon exchange changes the quark's color, and color is a conserved property, we can imagine that gluons, unlike quarks, carry both color and anti-color. For example, if a green quark emits a green-antiblue gluon, the quark must change its color to blue, because the total color must remain green. After emitting the gluon, the quark's blue color cancels out with the antiblue color of the gluon, and the remaining color is the quark's original green color.

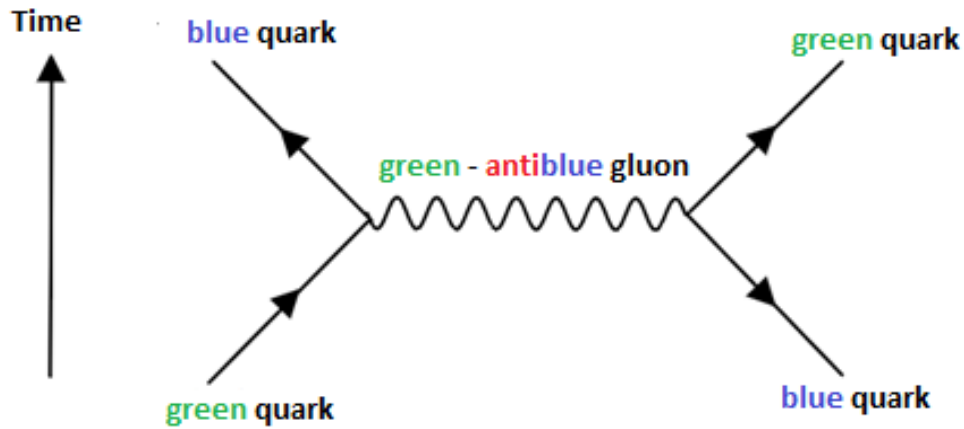

Figure 13 – Feynman diagram of blue and green quark interaction

By exchanging gluons, the two quarks in a hadron exchange colors in such a way that all quarks within the hadron are always differently colored, making the hadron "colorless."
The strength of the strong force by which quarks attract each other in hadrons depends on their mutual distance. However, interestingly, unlike all other fundamental forces, the strong force does not decrease its strength with increasing distance between quark pairs after reaching a certain distance limit - for distances greater than about the size of a hadron, it retains a value of about 10,000 N. [16] To attempt to separate a quark from a hadron, that quark would have to collide with another particle. By moving that quark away from the other quarks in the hadron, the potential energy of their interaction would increase, creating a quark-antiquark pair, or a meson. As a result, the potential energy would decrease again, and the quark that moved away would return to where it was before the collision. Therefore, adding energy to the hadron system cannot result in the liberation of individual quarks from the hadron; instead, new hadrons are formed in that case! This phenomenon, which has been observed in numerous experiments, is called "color confinement."[17]

## 8  Topics for further discussions

Below are some questions (and answers) related to this game, directly linked to the interactions of elementary particles, which can serve as topics for further conversation and discussion after playing the game. In this way, players (in workshops dedicated to the popularization of physics, or even in their own homes) can further refine and deepen the newly acquired knowledge of elementary particle interactions.

- ***The gravitational interaction in the game could be used only once. How would you interpret this rule in the real context of particle physics?***

(Gravitational interaction occurs between any particles with masses - between any quarks and leptons. However, the strength of the gravitational force between two particles is proportional to their masses (students can remember the equation), which makes it very weak for elementary particles. That „weakness" is reflected in the rule that allows using it only once during the game.)

• *According to the rules of the game, is there any interaction that can only occur between one type of elementary particles (quarks or leptons)?*

(Yes - it is the strong interaction, which can only occur between quarks but not leptons. Then we can discuss that, based on this fact, elementary particles are divided into two major groups - quarks (which can interact via the strong force) and leptons (which do not experience the strong force).)

• *What condition occurred in the game regarding the strong interaction, and how is this rule related to the real strong interaction among quarks?*

(In the game, strong interaction was possible between any two quarks that differ in color. This situation reflects the actual phenomenon of interaction between two quarks inside a baryon (where, for color neutrality of the hadron, there must always be three quarks of different colors). In such interaction, two quarks of different colors interact by exchanging gluons, resulting in the exchange of their colors (And then we can teach our students more about that.) Due to the short-range nature of the strong force, the game assumes that strong interaction can occur only among quarks within one baryon.)

• *What rule applied to the use of the weak force in the game regarding cards representing leptons? How could you connect this rule to the real weak interaction among leptons?*

(In the game, the weak force allowed the placement of leptons from the same generation on top of each other. This rule corresponds to the real weak interaction among leptons, which can only happen with leptons of the same generation but not between different generations. We then can refer to figure mentioned earlier, and tell more about the

principle of weak force – how it is acting to change the flavors of leptons within the same generation.)

• *According to experience and this game, what is common to all elementary particles (represented on the cards) that can interact through the electromagnetic force? How do elementary particles interact in this case?*

(Through electromagnetic interaction, any electrically charged particles, i.e., any quarks (represented in the game as u, s, and d quarks) or charged leptons (represented in the game as electrons and muons) can interact. Then we can list values of the electric charge for the particles depicted on the cards in the deck. In electromagnetic interaction, electrically charged elementary particles exchange photons, so maybe then we could discuss that with a little help from the Feynman diagram of electromagnetic interaction between two electrons.)

• *With the game experience, which particles interact the least with others? Why?*

(The particles that have the fewest interactions in the game are the electronic and muonic neutrinos. They cannot interact through the strong force (as they belong to the lepton group) or through the electromagnetic force (as they are electrically neutral). They can interact through the weak interaction, but since it is limited to leptons of the same generation, it is only possible to place an electronic neutrino on an electron and a muonic neutrino on a muon. The weak interaction of electronic, muonic, and tau neutrinos with the rest of the matter, which is why neutrinos are sometimes called "ghost particles," That can be further explored in a discussion about historical discoveries of particles and antiparticles.)

• *Based on the game, how would you explain neutrino oscillations? Why was their discovery very important?*

(According to the rules of the game, electronic and muonic neutrinos could be placed on top of each other, representing "neutrino oscillations." The inspiration for this game rule was derived from the physical phenomenon of neutrino oscillations, which refers to the periodic changes in neutrino flavor from one generation to another, revealing that neutrinos have mass. This discussion can also mention that Takaaki Kajita and Arthur B. McDonald were awarded the Nobel Prize in Physics in 2015 for the discovery of neutrino oscillations!)

• ***When is it possible for elementary particles to undergo annihilation? What must be the nature of particle pairs in that case?***

(Annihilation is possible between a particle and its antiparticle - we could discuss all the possibilities while constructing a table with all the rules for placing antiparticles on the central card in the game. This discussion can lead to learning more about antiparticles, their similarities and differences with respect to particles, as well as the laws that must be conserved in annihilation.)[5]

# 8 Acknowledgments

Thanks and credits are due to prof dr. sc. Dubravko Klabučar, a professor at Theoretical Physics Division of Particles and Fields at PMF Faculty of Science, Department of Physics :
As a mentor for my master's thesis, he  introduced me to the quark matter card games and provided exceptional support and assistance throughout the thesis preparation, and later on.  Additionaly, I would like to express my gratitude to prof. dr. Tamás Csörgő from MATE University in Hungary for the support and encouragement; together with prof. Klabučar, he encouraged me to me write this article, and to participate in the ISMD conference.
In conclusion, special thanks to all the aforementioned pioneers of card games with elementary particles, who have been a tremendous source of inspiration. Without their creativity and efforts, my ideas would not even exist.